\documentclass[a4paper,twocolumn,english,aps,prb,superscriptaddress,showpacs,amsmath,amssymb,byrevtex,floatfix]{revtex4}
\usepackage[T1]{fontenc}
\usepackage[latin1]{inputenc}
\usepackage{babel}
\usepackage{graphicx}
\usepackage[unicode=true,pdfusetitle,
 bookmarks=true,bookmarksnumbered=false,bookmarksopen=false,
 breaklinks=false,pdfborder={0 0 1},backref=section,colorlinks=false]
 {hyperref}

\makeatletter

\@ifundefined{textcolor}{}
{%
 \definecolor{BLACK}{gray}{0}
 \definecolor{WHITE}{gray}{1}
 \definecolor{RED}{rgb}{1,0,0}
 \definecolor{GREEN}{rgb}{0,1,0}
 \definecolor{BLUE}{rgb}{0,0,1}
 \definecolor{CYAN}{cmyk}{1,0,0,0}
 \definecolor{MAGENTA}{cmyk}{0,1,0,0}
 \definecolor{YELLOW}{cmyk}{0,0,1,0}
 }

\usepackage{dcolumn}\usepackage{bm}\usepackage{times}

\makeatother

\begin{document}

\title{Impact of edge shape on the functionalities of graphene-based single-molecule
electronics devices}

\author{D. Carrascal}

\affiliation{Departamento de Física, Universidad de Oviedo, 33007 Oviedo Spain}

\affiliation{Nanomaterials and Nanotechnology Research Center, CSIC - Universidad
de Oviedo, Spain}

\author{V. M. García-Suárez}

\affiliation{Departamento de Física, Universidad de Oviedo, 33007 Oviedo Spain}

\affiliation{Nanomaterials and Nanotechnology Research Center, CSIC - Universidad
de Oviedo, Spain}

\affiliation{Department of Physics, Lancaster University, Lancaster LA1 4YW, United
Kingdom}

\author{J. Ferrer}

\affiliation{Departamento de Física, Universidad de Oviedo, 33007 Oviedo Spain}

\affiliation{Nanomaterials and Nanotechnology Research Center, CSIC - Universidad
de Oviedo, Spain}

\affiliation{Department of Physics, Lancaster University, Lancaster LA1 4YW, United
Kingdom}

\date{\today}
\begin{abstract}
We present an ab-initio analysis of the impact of edge shape and graphene-molecule
anchor coupling on the electronic and transport functionalities of
graphene-based molecular electronics devices. We analyze how Fano-like
resonances, spin filtering and negative differential resistance effects
may or may not arise by modifying suitably the edge shapes and the
terminating groups of simple organic molecules. We show that the spin
filtering effect is a consequence of the magnetic behavior of zigzag-terminated
edges, which is enhanced by furnishing these with a wedge shape. The
negative differential resistance effect is originated by the presence
of two degenerate electronic states localized at each of the atoms
coupling the molecule to graphene which are strongly affected by a
bias voltage. The effect could thus be tailored by a suitable choice
of the molecule and contact atoms if edge shape could be controlled
with atomic precision.
\end{abstract}


\maketitle

\section{Introduction}

Graphene, which is a monolayer of carbon atoms arranged in a honeycomb
lattice, has unique properties excellently matching the requirements
for ideal electrodes: it is an extraordinary conductor, it is mechanically
extremely robust and chemically very stable, and it withstands electro-migration.
Recent advances in graphene nanoribbon fabrication and patterning\cite{Ponomarenko08,Gass08,Kosynkin09,Liying09}
shed weight on the plausibility of the use of graphene\cite{Wang09,Schedin07}
as a future nanoelectronics technology. The deployment of graphene
nanoelectronics should have inherent advantages, since the unique
two-dimensional nature of graphene should enable the fabrication and
visualization of stable nanometer-size devices of controlled shape.

The persistent inability to visualize and control the shape of bulky
noble meal electrodes at the atomic scale, as well as the positioning
of molecules inside the physical gap is hindering the blossoming of
molecular electronics as a mature science and its deployment as a
technology. Graphene's flat nature should enable the imaging and control
of the features of single-molecule junction with atomic precision
via scanning tunneling microscopy techniques. This could allow the
detailed characterization of the junctions by correlating the experimentally-determined
atomic arrangement with the electrical properties and with the results
of theoretical calculations. In addition, using molecules to bridge
graphene nanojunctions should furnish the array of expected applications
of graphene electronics with a whole suite of new functionalities\cite{Tour05,Bogani08}.

A crucial pending issue for the use of graphene nanoelectronics is
the difficulty in controlling the edges of graphene sheets. This could
influence not only the electronic properties of the graphene electrodes
but also change arbitrarily the coupling and switching functionality
of the molecules attached. However, controlled formation of sharp
zigzag and armchair edges has been achieved\cite{Dresselhaus-capulla}.
Furthermore, nanogap electrodes with gap-size below 10 nm have been
recently demonstrated, together with gating of single- or few-molecule
junctions\cite{Prins11}. Other approaches that could lead to the
controlled formation of nanogap graphene electrodes include atomic
force microscopy nanolithography of graphene\cite{He10} and atomically
precise fabrication of graphene nanoribbons by on-surface synthesis
methods\cite{Cai10}.

The tiny size of the currents involved in single- or few-molecule
electronic junctions render measuring them a difficult task. Various
solutions have been proposed to solve this problem, but until now
none of them have been thoroughly tested. The small size of the current is mainly a consequence of the mismatch between the on-site energy levels of the molecule and of the metallic electrodes, which reduces the coupling between them and therefore decreases the width of the transport resonances. A simple alternative would therefore consist of choosing electrodes with on-site levels that match to the molecular ones, since this would enhance the transmission through the system and improve the
transport properties. Graphene seems the perfect candidate because
carbon is the main ingredient in most molecular compounds.

We analyze here a series of molecular junctions whose electrodes are
graphene sheets terminated in a wedge geometry with either armchair,
zigzag or mixed edges as sketched in Figs. 1-2-3, which could mimic
the nanogap junctions fabricated in Ref. (\onlinecite{Prins11}).
We also mention that controlled fabrication of the mixed wedges shown
in Fig. 2 has been demonstrated in Ref. (\onlinecite{Dresselhaus-capulla}).
To simplify matters, the geometry of the proposed junctions is such
that the tips of the two wedges face opposite to each other. We have
simulated the wedges at two different separations, which are also
shown in Figs. 1-2-3. The different shapes and distances enable the
molecules to position themselves at the most favorable energy minima.
The tip-to-tip distance must be tailored to be of the order of, or
somewhat smaller than the length of the molecule, because a single
molecule of length $L$ can only bridge the two graphene wedges provided
that the two tips are positioned at a distance $d$ equal or slightly
smaller than $L$.

The aim of the article is to investigate the electronics and transport
properties of the above-defined graphene-wedge single-molecule junctions.
We find that even simple molecules such as Benzene-dithiolate (BDT)
or bipyridine (BPD) can lead to a large variety of non-trivial functionalities
provided a suitable wedge is chosen. Specifically, we show here that
some graphene-wedge BDT and BPD junctions show negative differential
resistance (NDR)\cite{Tour99}, spin-filtering (SP) behavior\cite{Park09}
and Fano resonances\cite{Finch09}. We mention that non-trivial functionalities such
as NDR\cite{Dragoman07,Cheraghchi08}, switching behavior\cite{Agapito07}
or spin filtering effects\cite{Koleini07,Yokoyama08} in graphene-based
atomic or molecular junctions have actually been predicted in the
recent past. We offer here a comprehensive analysis with differently-terminated
wedges, several junction gap lengths and two different molecules.
The large phase-space analyzed allows us to understand better how
this diverse functionalities could be tailored on demand if or when
graphene edge shapes could be controlled with atomic precision.


\begin{figure}
\includegraphics[width=0.65\columnwidth]{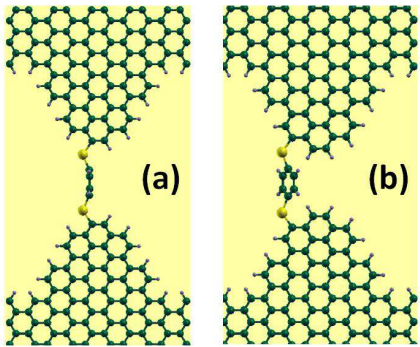} \caption{(Color online) Schematic view of an armchair wedge junction bridged
by a BDT molecule, where the tips at separated by (a) $9.4$ \AA{}~
and (b) $7.1$ \AA{}. The distance is measured between the last carbon
atoms at both wedges.}
\end{figure}

The SP effect can be seen when magnetic molecules join two paramagnetic
noble metal electrodes. This effect has been predicted to appear when
Mn$_{12}$ molecular magnets bridge gold electrones\cite{Park09}.
It originates because the nanojunction allows the passage of only
one of the two spin components from one side to the other. Spintronics
phenomena in graphene-based electronics were also proposed some time
ago\cite{Cohen06,Palacios07,Yazyev08} because zigzag edges either
unpassivated or passivated with a single hydrogen atom per carbon
atom are magnetic. 

The NDR effect is also sought after at the nanoscale due to its wide
variety of uses and applications such as in digital to analog converters,
oscillators, rectifiers and amplifiers \cite{Eisele98}. An efficient
NDR device is expected to display $I-V$ characteristics featuring a
sharp current peak at low bias voltages, immediately followed by a
low current minimum. Several mechanisms producing the NDR effect have
been proposed. In molecules between silicon electrodes it is generated
by the motion of resonances towards the silicon gap\cite{silicon};
this motion reduces the transmission through the system and decreases
the current. Another possible mechanism is a voltage-driven energy
mismatch of molecular levels which leads to the destruction of resonances\cite{Lyo89,Dalgleish06,Gar08,Chen07}.
The specific NDR effect demonstrated in this paper is originated by
the presence of pairs of electronic states placed close to the Fermi
level, and localized close to the molecule-graphene contacts. Each
of these two states moves in opposite energy directions under the
application of a bias voltage. This produces a reduction of the width
and height of the corresponding resonances in the transmission coefficients.
Therefore for a given voltage the current decreases, and the NDR effect
arises. In contrast, states that are localized in the middle of the
molecule are not so much affected by the bias voltage.

\begin{figure}
\includegraphics[width=0.9\columnwidth]{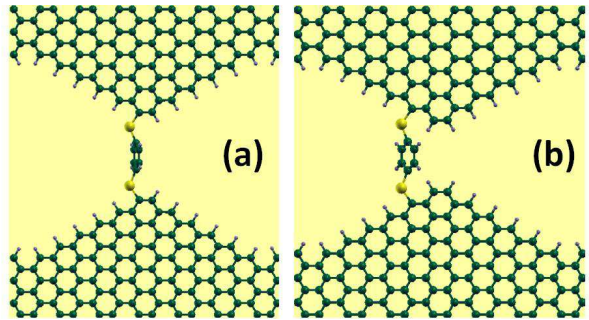} \caption{(Color online) Schematic view of a mixed zigzag-armchair wedge junction
bridged by a BDT molecule, where the tips at separated by (a) $9.5$
\AA{}~ and (b) $7.2$ \AA{}. The distance is measured between the
last carbon atoms at both wedges.}
\end{figure}

The layout of the article is as follows: details of the DFT theoretical
methods and simulations are presented in Section II. Transport through
BDT and BPD molecules is analyzed in Sections III and IV, respectively.
The article is closed with a discussion in Section V. A simple model
which explains the graphene-wedge NDR behavior found in the wedge
junctions presented here is included in the Appendix.

\begin{figure}
\includegraphics[width=0.7\columnwidth]{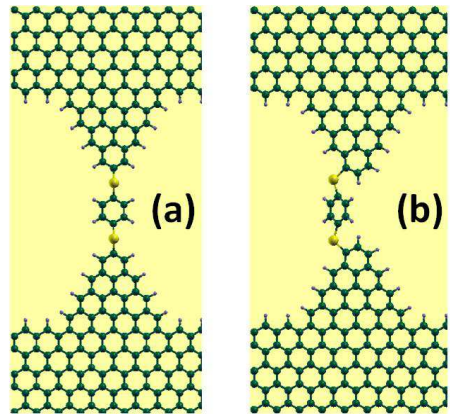} \caption{(Color online) Schematic view of a zigzag wedge junction bridged by
a BDT molecule, where the tips at separated by (a) $9.6$ \AA{}~
and (b) $8.2$ \AA{}. The distance is measured between the last carbon
atoms at both wedges.}
\end{figure}

\section{Theoretical method}

The ab-initio simulations have been performed with the density functional
theory (DFT) code SIESTA\cite{Soler02}, which uses norm-conserving
pseudopotentials to get rid of the core electrons, and pseudoatomic
orbitals in the basis set to span the valence states. We have used
in this case an optimized double-$\zeta$ basis set, which is enough
to describe accurately the graphene band structure. We have chosen
the exchange and correlation potential in the local density approximation
(LDA) as parametrized by Ceperley and Alder \cite{LDA}, but we expect
that our results should be robust enough to withstand the use of more
accurate functionals describing better the van der Waals interaction\cite{Roman09}.
The density, the Hamiltonian and the overlap matrix elements have
been calculated in a real space grid defined with a cutoff of 450
Ry. The structural relaxations of the junctions have been obtained
using a single $k$ point which is enough to converge those properties.
All forces have been relaxed up to a tolerance smaller than 0.001
eV/\AA{}. A correct analysis of the electronic structure including
the density of states (DOS) has required a 30$\times$30 $k$ grid.

\begin{figure}
\includegraphics[width=0.9\columnwidth]{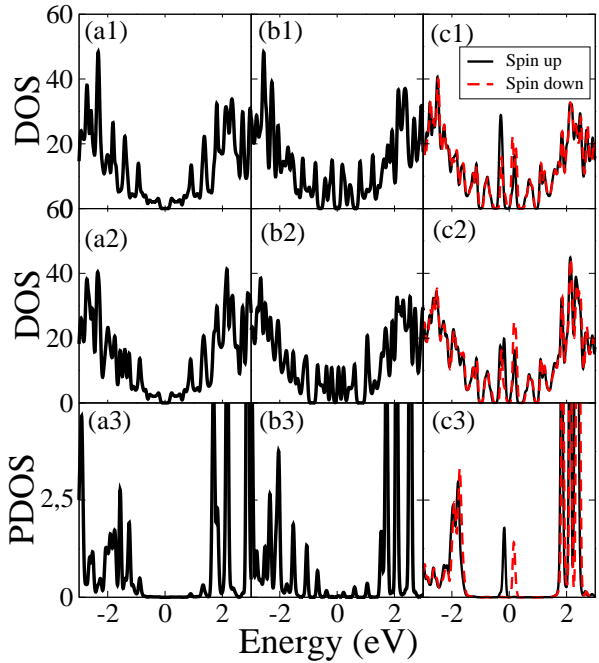} \caption{(Color online) (a1-b1-c1) Total DOS (in arbitrary units) of graphene
wedge junctions without a bridging molecule; (a2-b2-c2) Same, but
with a bridging BDT molecule. (a3-b3-c3) DOS (in arbitrary units)
of the BDT-bridged junctions, projected onto the molecule orbitals.
Left, center and right panels correspond to armchair, mixed and zigzag
wedges. Wedge-wedge distances correspond to the shortest-distance
cases shown in Figs. 1-2-3. Solid black and dashed red lines in panels
(c1), (c2) and (c3) correspond to up and down spins. }
\end{figure}

Portions of the simulation cells are sketched in Figs. 1-2-3. The
graphene sheets extend in the XZ-plane, with the wedges and nanojunctions
oriented along the Z-axis and the sheet width extended along the X-axis.
We have used periodic boundary conditions along the three spatial
directions, so that graphene sheets in neighboring simulation cells
are connected along the X and Z directions to avoid the presence of
additional edges that would distort the electronic structure. We have
measured sheet lengths counting the number $N$ of dimer/zigzag lines
for the armchair/zigzag direction, following the convention for graphene
nanoribbon unit cells. The simulated sheets have a width $N$ in the
range 10 to 20. Along the transport direction, they have 3 unit cells,
corresponding to $N=6$. We have then appended the wedges, whose edges
have been passivated with an hydrogen atom per carbon atom. 

The transport calculations have been carried out with the aid of our
non-equilibrium transport code SMEAGOL\cite{SMEAGOL}. The system
has been divided in three pieces: left lead, right lead and extended
molecule. The leads transport channels impinging from each lead onto
the molecule have been determined with a previous calculation for
each different lead. These leads calculations require simulating a
bulk unit cell with nearest neighbors coupling and periodic boundary
conditions. The lead's electronic structure must be converged using
a sizable number of $k$-points along the direction parallel
to the electronic transport. The lead's calculations serve to determine
the self-energies of the electrodes and to ensure that the electronic
structure at the two edges of the junction agree with those of a bulk
lead.

The extended molecule includes the central part of the junction (molecule
attached to graphene wedges) and also some layers of the graphene
leads to further ensure that the electronic structure at the edges
agrees with the lead's electronic structure. The atomic coordinates
must match those of the leads. Furthermore, basis functions and accuracy
parameters must also be consistent with those of the lead's calculations.
The code computes the electronic structure using the non-equilibrium
Green's functions formalism. We have used in the present calculation
10 $k$-points in the direction transverse to the electronic transport,
and have converged the density matrix down to a tolerance smaller
than 5x10$^{-5}$. We have extracted the transmission coefficients
of the junctions and its current at the end of each simulation. 

\begin{figure}
\includegraphics[width=1\columnwidth]{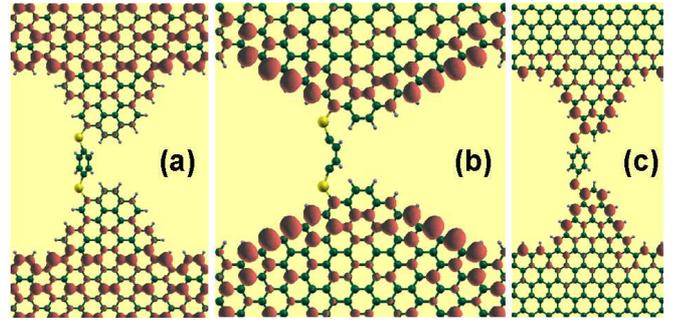} \caption{(Color online) Local Density of States integrated in an energy window
around the Fermi energy for (a) armchair, (b) mixed and (c) zigzag
wedges. For zigzag wedges we show only the majority spin component
of the charge.}
\end{figure}

\section{wedge-BDT-wedge junctions}

We describe first the details of the force relaxation simulations
of the wedge junctions bridged by BDT molecules. We have simulated
first fully passivated wedges. For each given simulation, we have
placed initially a single Benzene-dithiolated molecule close to the
two tips, and then have relaxed the forces. Most of these simulations
end with a fully developed junction. Both the graphene wedges and
the molecules are slightly bent or deformed at the end of the simulations
to accommodate for their chemical bonding. Further, we have performed
simulations that include two molecules to check the rate of single
versus double molecule junctions. Indeed, both single or double junction
formation are achieved when the two molecules are initially placed
at opposite sides of the tips. However, we have found that if the
molecules are placed initially not too far away from each other, the
sulfur atoms at the two molecules rather bind to each other and no
junction is developed. We are led to conclude that junction formation
is highly implausible for fully passivated wedges and thiolated molecules,
since the molecules bind to each other time before reaching the junction
area. To fix this problem, we have substituted the thiolate by thiol
end-groups at the sides of the molecules, We have checked that thiol-capped
molecules do not bind to each other, but their reactivity is so reduced
that they do not bind to the graphene edges either. We have indeed
been unable to achieve the formation of a single junction in any of
the simulations carried out.

\begin{figure}
\includegraphics[width=0.9\columnwidth]{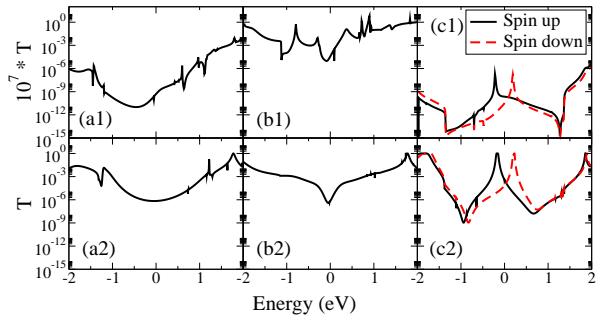} \caption{(Color online) (a1-b1-c1) Transmission coefficients of graphene wedge
junctions without a bridging molecule multiplied by$10^{7}$; (a2-b2-c2)
Same, but with a bridging BDT molecule. Left, center and right panels
correspond to armchair, mixed and zigzag wedges. Wedge-wedge distances
correspond to the shortest-distance cases shown in Figs. 1-2-3. Solid
black and dashed red lines in panels (c1) and (c2) correspond to up
and down spins. }
\end{figure}

We have decided as a consequence to passivate the wedges only partially,
so that one or several of the passivating hydrogen atoms in the vicinity
of the tip area are missing, and to cap the molecules with thiol,
instead of thiolated, end-groups to avoid the chemical bonding between
any two of them. We have found that the edge areas containing an unpassivated
carbon and its nearest neighbor carbon atoms are extremely reactive.
When the molecules reaches the unpassivated area, their sulfur atoms
are stripped off their hydrogen terminations. Molecule-wedge chemical
bonds are hence formed, where the sulfur atoms are bound to the previously
unpassivated carbon atoms, and the stripped hydrogens are attached
to other nearby carbon atoms at the graphene edge. Further, the hydrogen
affinity of the edges is so large that in some cases the inner carbon
atoms of some of the molecules are also stripped off their hydrogen
atoms, which end up attached to other previously unpassivated edge
carbon atoms. We have performed simulations with unpassivated areas
at both wedges. We have found that junctions are made only when the
distance $d$ between the unpassivated carbons is approximately equal
to the length $L$ of the molecule. Those simulations where $d$ is
larger or considerably shorter than $L$ end up with the molecule
being attached to only one of the passivated areas, but not to the
other. In contrast, those simulations where $L$ is only slightly
larger than $d$ most likely end up with the formation of the junction.
To this end, both the molecule and the tips bend and distort slightly,
to make room for the chemical bond. Typical final configurations are
shown in Figs. 1-2-3. For armchair wedges the molecule is bound to
carbon atoms at the side of the wedge tip and is oriented roughly
perpendicular to the graphene planes. This effect is due to the particular
directionality of the C-S bond, which give non-trivial structural
configurations. We have found a similar effect for the mixed wedges.
In contrast, zigzag wedges are terminated in a single carbon atom.
In this case, BDT molecules find it easier to orient roughly parallel
to the graphene layers.

\begin{figure}
\includegraphics[width=0.9\columnwidth]{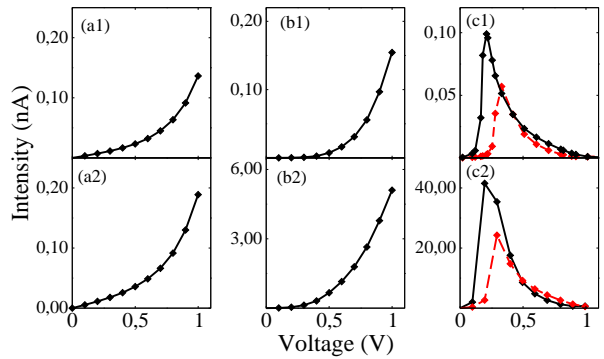} \caption{(Color online) $I-V$ characteristics of the system composed of wedge-BDT-wedge
junctions for (a) armchair, (b) mixed and (c) zigzag edges. Top/bottom
panels correspond to longest/shortest distances between the graphene
edges shown in Figs. 1-2-3. Solid black and dashed red lines in panels
(c1) and (c2) correspond to up and down spins.}
\end{figure}

We have addressed the mechanical stability of the junctions at room
temperature by carrying out Molecular Dynamics simulations using a
Nose thermostat with target temperature of 300 K. The temperature
profile of a typical run then oscillate about the target temperature
as a function of the Molecular Dynamics steps, with temperature peaks
as large as 500 K for some of the steps. We have found that the molecule
and carbon atoms at the junction area undergo collective motions as
the vibrational modes of the junction are activated. But even with
these strong vibrations, the integrity of the junction is maintained
throughout all the steps of all the simulations carried out. These
simulations hence indicate that graphene-wedge molecular junctions
could be stable at room temperature.

We analyze now the electronic structure of the three wedge junction
types shown in Figs. 1-2-3. The top panels in Fig. 4 show the total
density of states (DOS) of the junctions without bridging molecules,
which sheds a tentative idea of how the leads conduction channels
are distributed in energy. Note that the DOS shape of armchair wedges
(Fig. 4(a1)) resembles closely bulk graphene's DOS, as is the case
of large-width graphene nanoribbons\cite{Dresselhaus-nr}. The peaks
are due to the finite number of $k$-points used. In addition these junctions
show a small gap about the Fermi energy $E_{\mathrm{F}}$. Mixed and
zigzag wedges show additional peaks about $E_{\mathrm{F}}$ associated
with edge states. These edge states hybridize with BDT's frontier
orbitals giving rise to a finite DOS projected on the molecule orbitals
(PDOS) as shown in Figs. 4(b3-c3). In addition, zigzag wedges are
spin-polarized. This spin polarization is also transmitted to the
BDT molecule, as demonstrated by the peaks in the PDOS shown in Fig.
4(c3). In contrast, armchair and mixed wedges are non-magnetic.

\begin{figure}
\includegraphics[width=0.6\columnwidth]{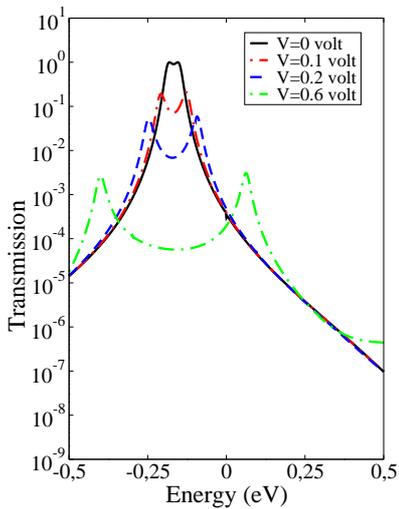} 
\caption{\label{BDT.Zigzag_sharp.TRC_vs V}(Color online) Spin-polarized energy-dependent
transmission of a BDT zigzag junction, plotted for voltages increasing
from 0 to 0.6 volt in a logarithmic scale.}
\end{figure}

To learn how these edge states are distributed spatially, we have
plotted the wedge Local Density of States (LDOS), integrated in a
window from $-0.5$ to $+0.5$ eV about the Fermi energy. We have
found that charge is driven away from the edges for armchair wedges,
as is shown in Fig, 5(a). In contrast, mixed wedges have large edge
segments with zigzag arrangements, which show accumulation of charge.
The tip of these wedges however shows an armchair arrangement; consequently,
charge is driven away from the wedge tip, as is shown in Fig. 5(b).
Zigzag wedges show spin-polarized edge states all the way up to the
wedge tip, except for the last carbon atom, as can be seen in the
LDOS plotted in Fig. 5(c).

The junctions transmission coefficients $T(E)$ are shown in the three
bottom panels in Fig. 6. The top panels show the transmission coefficients
multiplied by $10^{7}$ for similar junctions, where the BDT bridging
molecules have been replaced by passivating hydrogen atoms. In addition
the gap length in this second set of junctions has been adjusted a
little to allow for a tiny but finite overlap between those terminating
hydrogen atoms, so that transmission coefficient would be non-zero.
Notice that the behavior of the top panels roughly matches with that
of the equivalent bottom panels, signaling that the functionalities
of these devices are brought about by the specific wedge shape in
each case. This was of course expected given the known non-functional
behavior of BDT molecules when contacted by noble metal electrodes.
The transmission of armchair wedges is quite featureless and small,
while for mixed wedges the transmission is small but shows a marked
dip pinned at the Fermi energy. Zigzag wedges are much more interesting.
Here, sharp transmission peaks around the Fermi energy are apparent.
$T$ shows only one spin polarized peak on each side of the Fermi
level which is correlated with the peaks in the DOS curves shown in
the right panels in Fig.4. We expect that the sharp peaks should give
rise to much higher values of the current in the $I-V$ characteristics
than for the former two types of junctions. Furthermore, spin-filtering
phenomena should appear.

\begin{figure}
\includegraphics[width=1\columnwidth]{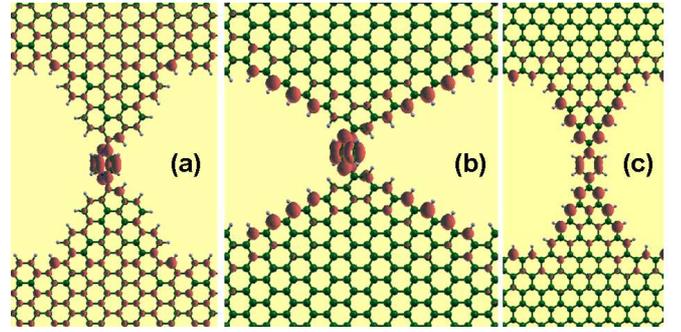} \caption{(Color online) Local Density of States of BPD junctions integrated
in an energy window around the Fermi energy for (a) armchair, (b)
mixed and (c) zigzag wedges. For zigzag wedges we show only the majority
spin component of the charge.}
\end{figure}

We calculate the current by integrating the bias dependent transmission
coefficients $T(E,V)$ in an energy window given by the voltage bias.
The $I-V$ characteristics for armchair junctions show ohmic-like
behavior at small voltages as shown in Figs. 7(a1-a2). The absence
of transmission resonances close to the Fermi level is manifested
in the small values of the current. This behavior can also be understood
by inspecting the LDOS shown in Fig. 5(a), where it is apparent that
the charge within the graphene wedges accumulates far away from the
junction. Mixed junctions show a semiconducting shape, where a gap
is apparent followed by a steep rise in the conductance. Further,
the current changes by more than one order of magnitude when approaching
the two edges from 9.5 \AA{} to 7.2 \AA{}. This can again be understood
by inspecting Fig. 5(b), where it is clear that the BDT molecule attaches
to atoms having a finite LDOS for the closer distance (corresponding
to Fig. 2(b)), but not when the edges are separated by 9.5 \AA{}~
which corresponds to Fig. 2(a). For zigzag junctions, the sharp spin
polarized transmission peaks at low energies result in a spin filtering
effect which we show in Figs. 7(c1-c2). These $I-V$ characteristics
also show a strong NDR effect, whose origin can be traced to a shift
in energy and a change in shape of the spin-polarized transmission
peaks shown in Fig. 6(c1-c2) as the bias voltage increases, as we
will show below. Reducing the distance between the two edges from
9.6 \AA{}~ to 8.2 \AA{}~ yields a change in the current of more
than two order of magnitude. To understand this change one must again
look at Fig. 4(c) and realize that when the molecule attaches right
at the wedge tip (Fig. 3(a)), there is no charge available for transport
at low voltages, while if the molecule attaches to the neighboring
carbon atom (Fig. 3(b)), then the edge state hybridizes strongly with
the sulfur atom as is shown in Fig. 5(c).

\begin{figure}
\includegraphics[width=0.9\columnwidth]{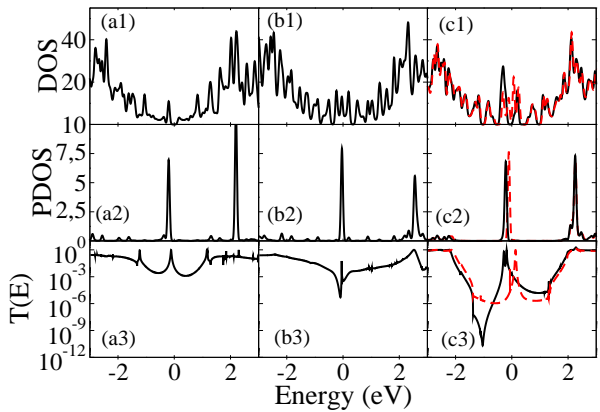} 
\caption{(Color online) (a1-b1-c1) Total DOS. (a2-b2-c2) DOS projected on the
molecule atoms. (a3-b3-c3) transmission coefficients of wedge-BPD-wedge
junctions. Left, middle and right panels correspond to armchair, mixed
and zigzag wedges. Solid black and dashed red lines in panels (c1),
(c2) and (c3) correspond to up and down spins.}
\end{figure}

To understand better the origin of the NDR effect in zigzag BDT junctions
we plot in Fig. 8 the transmission coefficients $T(E,V)$ as a function
of energy $E$ for several voltages $V$ increasing from 0 to
0.6 volt. Notice that the height of the peaks decreases and splits
in two as the voltage is increased. To understand this behavior we
focus now on the molecular orbitals associated with the transmission
peaks. These are localized mainly at the sulfur atoms placed at the
two edges of the molecule. They are responsible for the coupling between
molecule and electrodes. These orbitals give rise to bonding and antibonding
molecular orbitals across the molecule. It turns out that the impact of a voltage bias
on these frontier molecular is strong and non-trivial. When $V$ increases
the degree of localization at the coupling atoms increases, they become
asymmetrically coupled to the electrodes and their energy moves in
opposite energy directions. The evolution of the bonding and antibonding
states produced by the above molecular orbitals with the bias voltage
is the main mechanism triggering the NDR behavior in these junctions.
The mechanism however is generic and could be applied to any other
junction where frontier molecular orbitals are located close to the
Fermi energy. We therefore suggest that a necessary condition for
the appearance of NDR in small molecules coupled to metallic electrodes
is the existence of these side molecular orbitals. A simple model
that captures the essential features of the NDR effect consists of
two metallic monoatomic chains having localized states at the edges
which interact across a vacuum region, as we explain in the Appendix.
The more localized the states and the less interaction between them
the lower the voltage where the NDR peak appears and the higher the value of the peak to valley ratio but the lower also
the value of the current.
The key features of an efficient NDR device are a low value of the
peak voltage and a high peak to valley value of the current. These
can be tuned by balancing the degree of localization of the edge states
and their coupling across the vacuum region.

\begin{figure}
\includegraphics[width=0.98\columnwidth]{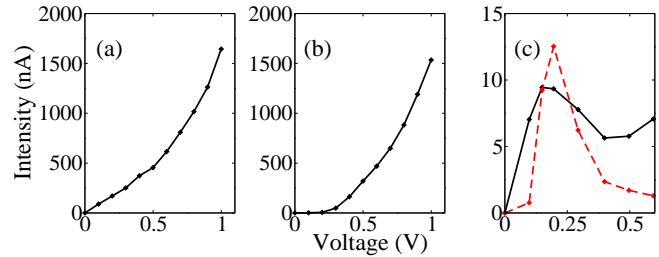} 
\caption{(Color online) Current as a function of bias voltage for BPD junctions
with (a) armchair (b) mixed and (c) zigzag wedges.}
\end{figure}

\section{Wedge-BPD-Wedge junctions}

To test for the impact of the contact atoms on the transport properties,
we have carried out simulations where the bridging BDT molecule in
the junctions shown in Figs. 1(a), 2(a) and 3(a) has been replaced
by a BPD molecule. This means that the new set of calculations has
been performed only for intermediate junction gap lengths of about
9 \AA{}, and with the geometries shown in Fig. 9. The total DOS and
the DOS projected onto BPD's nitrogen and carbon atoms are shown in
Figs. 10(a1-b1-c1) and 10(a2-b2-c2), respectively. Notice that the
DOS shows now a sharp peak close to the Fermi level for the three
wedge junctions. To check for the spatial location of the state associated
to that peak we have plotted in Fig. 9 the LDOS integrated in a small
window in energy which only retains the peak. The figure demonstrates
that the state corresponds to a molecular orbital extended throughout
the whole molecule now. The corresponding transmission peak or Fano-like
resonances, shown in the bottom panels in Fig. 10, are pinned at the
Fermi energy and should lead to a dramatic increase in the current
of armchair and mixed wedge junctions, as compared to their equivalent
BDT junctions. The $I-V$ curves of these wedge junctions are shown
in Fig. 11. Notice indeed that the current of the armchair and mixed
wedge junctions increases by several orders of magnitude, although
both $I-V$ curves retain the ohmic- and semiconducting-like behaviors displayed in BDT
junctions. BPD armchair and mixed junctions do not show NDR behavior
because the new molecular state associated with the PDOS peak is localized
inside the molecule and not at its edges, and is therefore not strongly
affected by the bias voltage. Zigzag BPD junctions also show an increase
of the current of two orders of magnitude compared to Fig. 7 (c1) although the effect is not
so dramatic as for armchair and mixed junctions, see Fig. 11(c). To
understand this we compare in Figs. 9(a), (b) and (c) the amount
of charge associated with the molecular state, which is clearly smaller
for the zigzag junction. The SP and NDR behaviors of the zigzag BPD
junction shown in Fig. 11(c) are not so marked as for the BDT molecule
as well. The reason behind the reduction of the two effects can also
be traced to the new molecular state responsible for the current increase.
The point is that this molecular orbital is not so affected by the
graphene edge states as the sulfur atoms in BDT junctions were.

Fig. 12 shows the behavior of the energy-dependent transmission coefficient
of a BPD mixed junction as the voltage is ramped up from 0 volt. Notice
how a Fano-like resonance develops such that a dip feature emerges
as $V$ increases. Further, notice that the peak and the deep move
in opposite energy directions to make the resonance wider. Because
the peak moves to lower negative energy values when $V$ increases,
a large fraction of its weight remains out of the integration window.
This is the reason why the $I-V$ characteristics of mixed junctions
show a semiconducting shape. The voltage dependence of the transmission
coefficients of zigzag junctions is shown in Fig. 13. Here the split
frontier orbitals are non-degenerate even at zero voltage. Ramping
up the bias further splits the peaks. However the position and height
of the dominant peak does not change much. This peak corresponds to
the molecular orbital centered within the BPD molecules, which is
not too strongly affected by the bias. In contrast the position and
height of the other peak are severely affected by the voltage. This
NDR effect is not so marked as for the zigzag BDT junction because
here it results of a trade-off of the evolution of the different transmission
peaks.

\begin{figure}
\includegraphics[width=0.9\columnwidth]{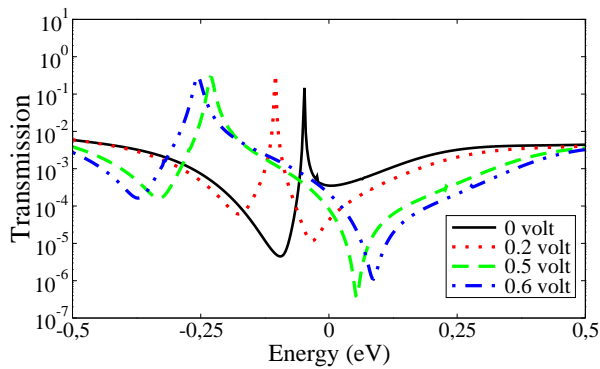} 
\caption{(Color online) Energy-dependence of the transmission coefficient of
a BPD mixed junction for voltage biases ramped from 0 to 0.6 volt.}
\end{figure}

\section{Conclusions}

This article presents an analysis of the anchoring of benzene-dithiolate
and bipyridine molecules to a variety of graphene wedge electrodes.
It also presents results of the electronic and transport properties
of the ensuing single-molecule junctions. The article shows how the
diverse electronic structure of the graphene edges is transmitted
into a rich variety of transport phenomena, which include spin-polarized
transport, negative differential resistance behavior and Fano-like
resonances. On the positive side, the wealth of results obtained for
simple non-functional molecules indicates that graphene-wedge single
molecule electronics could be exploited as a fruitful playground for
new functionalities provided that the shape of the edges could be
controlled. On the negative side, such a variety of behaviors should
lead to strong tribological effects and large variability in the electrical
response of these junctions if edge shape morphology is not controlled
at the atomic scale.

\begin{figure}
\includegraphics[width=0.9\columnwidth]{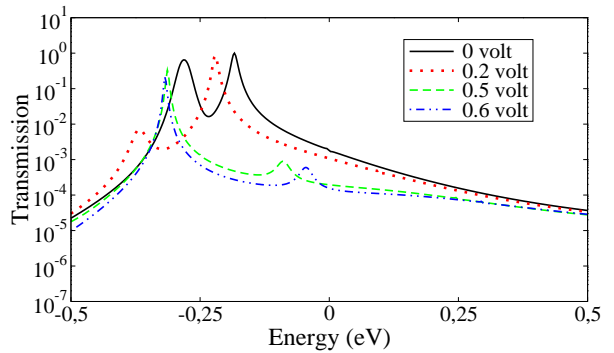} 
\caption{\label{Pyridine.Zigzag_sharp.TRC_vs V}(Color online) Energy-dependence
of the spin-up transmission coefficient of a BPD zigzag junction for
voltage biases ramped from 0 to 0.6 volt. }
\end{figure}

\acknowledgments This work was supported by the the Spanish Ministry
of Education and Science (project FIS2009-07081) and the Marie Curie
network NanoCTM. VMGS thanks the Spanish Ministerio de Ciencia e Innovación
for a Ramón y Cajal fellowship (RYC-2010-06053). JF acknowledges discussions
with C. J. Lambert, M. Calame, F. Prins and R. Fasel.

\appendix

\section{Simple model for the NDR effect in graphene-wedge junctions.}

The main mechanism behind the NDR effect in molecules between graphene
sheets and in many other molecular junctions is produced by the localization
of states at the contacts which are weakly coupled through
the molecular backbone and have energies close to the Fermi level.
These states separate as the bias increases, which decreases the transmission
through the junction near the Fermi level and reduces
the current for biases larger than a certain bias (threshold voltage).
As a consequence, the increase of the integration voltage window can
not compensate the reduction of the transmission and the total current
is reduced from the maximum obtained at lower absolute voltages.

\begin{figure}
\includegraphics[width=0.9\columnwidth]{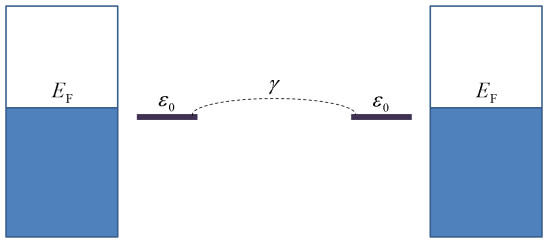} \caption{\label{Two_levels}(Color online) Two level system used to simulate NDR mechanisms produced by the localization of states near the contacts.}
\end{figure}

A simple picture that can grasp this effect can be elaborated with a two level system such as the one shown in Fig. (\ref{Two_levels}). Both
levels are located at the contacts and have an energy equal to $\varepsilon_{0}$,
which in this case it a bit below the Fermi level of the leads, $E_{\mathrm{F}}$,
but could also be a bit above. The coupling between the levels, $\gamma$,
greatly influences the shape of the transmission. If $\gamma$ is
big enough both levels interact strongly and give rise to bonding
and antibonding states separated $2\gamma$. Since the system is
symmetric, these levels produce Breit-Wigner resonances of height
equal to one (or a integer multiple of one in case of degenerate states).
When $\gamma$ decreases both levels merge and the bonding-antibonding
system transforms into a system with two asymmetric levels located
at the same energy. As a consequence the Breit-Wigner resonances
merge into a single resonance which corresponds to the transmission of two asymmetrically
coupled levels and whose width decreases as $\gamma$
is further reduced. A example of system with
contact states relatively well coupled through the molecule is
the benzene-dithiolate molecule between gold leads, whose transmission shows a broad HOMO resonance of height almost equal to one produced by states located on the sulfurs. An example of system with contact states weakly coupled through the molecule is e.g. any alkane-dithiolate molecule between gold leads, whose sulphur states do not interact across the molecular backbone and give rise to transmission resonances of heights much smaller than one.

It is important that such levels are relatively near the Fermi
level, localized on the contacts and coupled through the molecule, so that they produce sharp resonances. Under such conditions the effect of the bias is enhanced due to the quick misalignment of the resonances, which strongly reduces the transmission and the current and gives rise to clear NDR effects. The sharper the
resonance and the closer it is to the Fermi level the better the
NDR parameters. However, if the levels strongly hybridize with the electrodes the transmission features are much broader and the NDR develops at much higher voltages or does not develop at all. On the other hand, if sharp resonances are located away from the Fermi level the resulting NDR peak broadens and moves to higher voltages since the effect of the reduction of the height of the resonances is smaller away from the position of resonance peak. Something similar happens if the resonances are close to the Fermi level but are broad instead. In that case the height of the transmission features decreases more slowly and the NDR fades again.

\begin{figure}
\includegraphics[width=0.9\columnwidth]{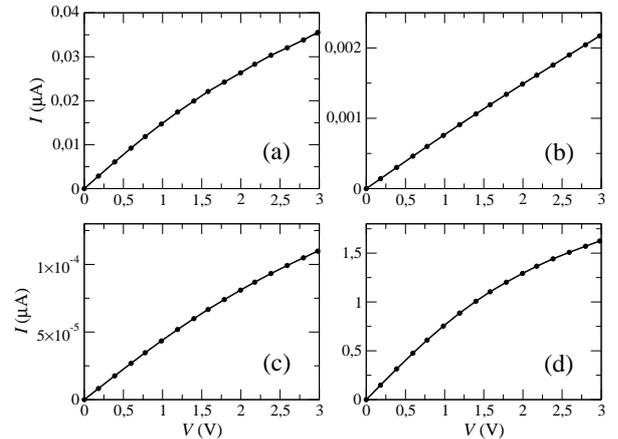} \caption{\label{Curr_gap.C_El.Relax}Current as a function of voltage for junctions made of carbon (a), nitrogen (b), oxygen (c) and sulfur (d) atoms connected to perfect atomic carbon chains
and separated by a vacuum region of 3.70 \AA{}\.{T}he atomic coordinates
of the contact atoms have been relaxed.}
\end{figure}

\begin{figure}
\includegraphics[width=0.9\columnwidth]{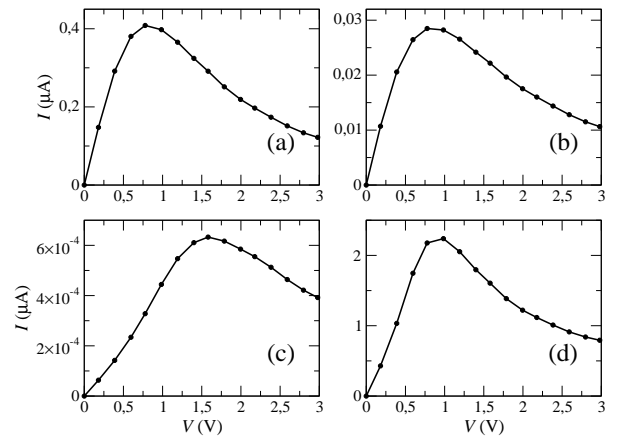} \caption{\label{Curr_gap.C_El.Relax+0.5}Current as a function of voltage for
junctions made of carbon (a), nitrogen (b), oxygen (c) and
sulfur (d) atoms connected to perfect atomic carbon chains and separated by a vacuum region of 3.70 \AA{}\.{T}he distance
between the atoms and the chain has been artificially increased 0.50
\AA{}\ from the relaxed distance.}
\end{figure}

It is possible to conclude then that in all cases with localized states on the coupling region the development and shape of the NDR depend on the localization of such states and
their interaction across the molecule. This last variable can be simplified and the system can be transformed into a simple case of two coupling
atoms with a vacuum region in between, which would mimic an ideal low-conductive molecule. This model has basically two variables, the distance between the coupling atoms and the electrodes and the length of the vacuum region between them. The first variable determines the localization of the states, which is directly related to the width of the transmission resonances, and the second variable determines the height and shape of the resonances (i.e. resonances of height equal to one when the distance is small and the atoms are well coupled, and resonances of smaller height when the distance is large and the atoms are weakly coupled). A third variable could be the position of the contact states, determined by the type of atom. 

\begin{figure}
\includegraphics[width=0.9\columnwidth]{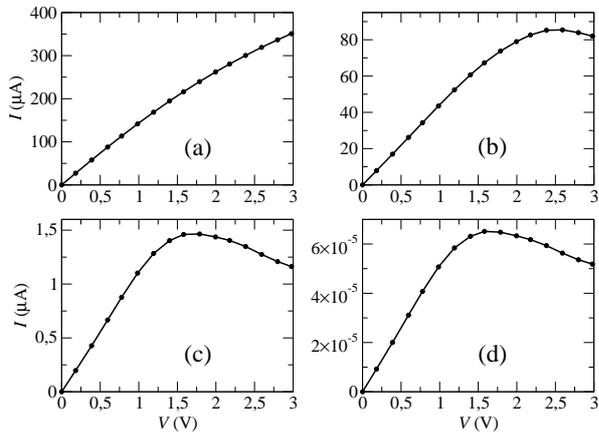} \caption{\label{Curr_gap.C_S.No_relax.d-d}Current as a function of voltage
for junctions made of sulfur atoms connected to perfect atomic carbon
chains, a bit stretched from the relaxed distance and separated by
a vacuum region of 1.74 \AA{}\ (a), 2.74 \AA{}\ (b), 3.74 \AA{}\ (c)
and 5.04 \AA{}\ (d).}
\end{figure}

We simulated this system with various types of light elements typically used to couple molecules to metallic electrodes(C, N, O and S) bonded to perfect atomic chains made of carbon (1.28 \AA{}\ of separation between atoms), which would mimic perfect metallic electrodes. We chose a initial distance of 3.70 \AA{} between the coupling atoms,
which couples them but not very strongly. We used the LDA approximation for the exchange and correlation potential,
a real space mesh cutoff of 200 Ry and a double-zeta polarized (DZP)
basis set. Even though some of the cases would be magnetic, we did not include the possibility of spin-polarization to further simplify the system and focus only on
the NDR effect. We relaxed the position of the coupling atoms until
the forces were smaller than 0.05 eV/\AA{}, which gave distances to
the first atom of the carbon chain of 1.31 \AA{}, 1.19 \AA{}, 1.17
\AA{} and 1.57 for C, N, O and S, respectively. 

The results obtained with the initial
parameters are shown in Fig. (\ref{Curr_gap.C_El.Relax}). As can
be seen all currents are almost Ohmic and do not show any sign of
NDR. This is because at the relaxed distances all coupling atoms
hybridize strongly with the carbon chain and do not produce resonances
in the transmission coefficients but plateaus of almost constant transmission,
like the one shown in Fig (\ref{Trans_gap.C_N.Relax+d}) (a). This
case however is not typical in molecular electronics systems,
where the coupling to the electrodes is not as strong and the states
on the coupling atoms are more localized (unless the molecules are very small, diatomic-like\cite{Ferrer09}). To further increase the localization
we artificially separate the coupling atom 0.5 \AA{}\ from the carbon
chain. This change produces the results shown in Fig. (\ref{Curr_gap.C_El.Relax+0.5}).
A clear NDR behavior appears now in all cases. The localization produces transmissions that
have a resonance-like shape. The height of such resonances is reduced
and the peak is divided in two smaller peaks (corresponding to the coupling states
on each side) as the voltage increases. 

When the coupling atom is carbon the atomic states closer to the Fermi level, which coincide with the main
transmission peaks, are located a bit above $E_{\mathrm{F}}$. In case of N and O they appear at lower energies (very close to the Fermi level, and below it, respectively), following the expected evolution due to the increasing nuclear attraction from C to O. In case of S the states are a bit above those of O, which is also expected due to the lower electronegativity of S compared
to O. Notice also the current is highest when the coupling atom is S and decreases
from C to O. This is due to the nuclear attraction again, which increases the atomic localization of the states and therefore decreases their interaction across the vacuum
gap as the atomic number increases along the same row of the periodic
table or decreases along the same column.

\begin{figure}
\includegraphics[width=0.9\columnwidth]{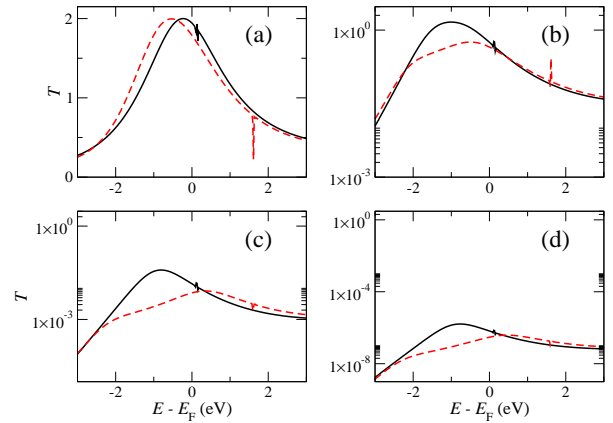} \caption{\label{Trans_gap.C_S.No_relax.d-d}Transmission of junctions made
of sulfur atoms connected to perfect atomic carbon chains, a bit
stretched from the relaxed distance and separated by a vacuum region
of 1.74 \AA{}\ (a), 2.74 \AA{}\ (b), 3.74 \AA{}\ (c) and 5.04 \AA{}\ (d).
Continuous and dashed lines correspond to the transmission at 0 V
and 3 V, respectively. Notice that in panel (a) the scale is linear.}
\end{figure}

The element that gives the larger coupling across the vacuum gap is
sulfur. We use then this element to study the effect of changing the
distance between the contact atoms, maintaining the distance between
S and the carbon chain equal to the relaxed distance plus 0.3 \AA{} to produce some localization.
The results are shown in Fig. (16). As can be seen the $I-V$ characteristics
evolve from almost ohmic to NDR-like. This evolution can be explained
by taking into account the transmission, shown in Fig. (\ref{Trans_gap.C_S.No_relax.d-d}).
When the atoms are very close the transmission around the Fermi level
has a Breit-Wigner resonance of height equal to two produced by two degenerate states corresponding to the p orbitals perpendicular
to the transport direction. In such situation the delocalized molecular states screen
the effect of the bias voltage and the resonance moves only slightly to
lower energies due to charge transfer\cite{Gar08}. If the distance increases, however, both states uncouple and separate. This allows them to follow the chemical potential of each electrode
at finite biases. In such a case the net effect on the transmission produced by the bias is a drastic reduction
of the resonance height and its further separation in
two satellites.

\begin{figure}
\includegraphics[width=0.9\columnwidth]{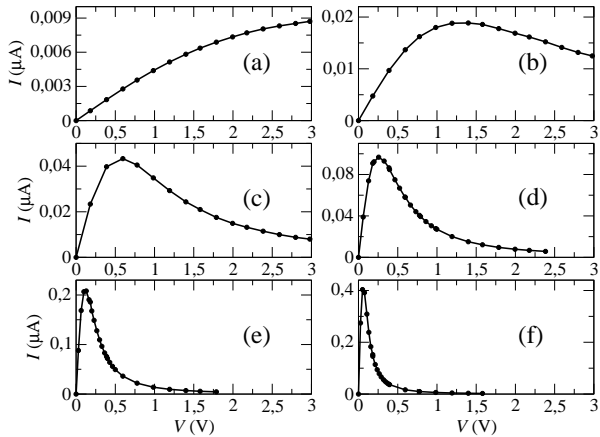} \caption{\label{Curr_gap.C_N.Relax+d}Current as a function of voltage for
junctions made of nitrogen atoms connected to perfect atomic carbon
chains, separated by a vacuum region of 3.70 \AA{}\ and stretched
from the relaxed distances 0.20 \AA{}\ (a), 0.40 \AA{}\ (b), 0.60
\AA{}\ (c), 0.80 \AA{}\ (d), 1.00 \AA{}\ and 1.20 \AA{}.}
\end{figure}

The factor that dramatically improves the quality to NDR is the localization of
the coupling states, which can be increased by varying the distance
between the coupling atom and the carbon chain: the larger the distance
the more localized the atomic states and viceversa. We used N to study this effect, since it has the closest states to the Fermi
level and can produce the clearest NDR signals. The results are shown
in Fig. (\ref{Curr_gap.C_N.Relax+d}). As can be seen, the height of the NDR peak increases and the voltage at which it appears decreases as a function of distance. This evolution can be easily understood by looking at the transmission as a function of bias, shown in Fig. (\ref{Trans_gap.C_N.Relax+d}) at 0 V and 1.6 V. When the distance is
small the large hybridization with the carbon chain does not produce
any resonance and therefore the transmission does not change with
bias. As the distance increases a transmission peak starts to develop.
The larger the distance the higher and sharper the peak and the bigger the effect
of the bias voltage, which splits the peak in two smaller satellites. This can be explained by taking into account that at very large distances the coupling
between the contact atoms is larger than the coupling to the chain;
this coupling effectively produces molecular states in the middle
of the junction that generate very sharp resonances in the transmission.
When the voltage increase the states of each atom follow the
chemical potential of their electrode and, if the screening is not very big, the system transforms into a system of two asymmetric states, each of which produce a smaller transmission peak.

\begin{figure}
\includegraphics[width=0.9\columnwidth]{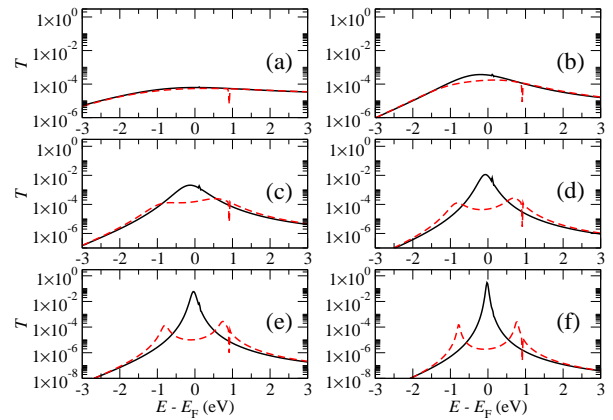} \caption{\label{Trans_gap.C_N.Relax+d}Transmission of junctions made of nitrogen
atoms connected to perfect atomic carbon chains, separated by a vacuum
region of 3.70 \AA{}\ and stretched from the relaxed distances 0.20
\AA{}\ (a), 0.40 \AA{}\ (b), 0.60 \AA{}\ (c), 0.80 \AA{}\ (d),
1.00 \AA{}\ and 1.20 \AA{}. Continuous and dashed lines correspond
to the transmission at 0 V and 1.6 V, respectively.}
\end{figure}

In summary, a very simple model that takes only into account the coupling of the contact atoms to the electrodes and the coupling between them can be used to describe fairly well a NDR mechanism that is
expected to appear in many molecular junctions. In reality the
molecular backbone would change the hybridization and position of
the contact states but the conclusions reached here would still be
valid, i.e. the more localized and closer to the Fermi
level the contact states, the better the NDR.

{}

\begin{thebibliography}{References}
\bibitem{Ponomarenko08} L. A. Ponomarenko, F. Schedin, M. I. Katsnelson,
R. Yang, E. W. Hill, K. S. Novoselov and A. K. Geim, 
Science \textbf{320}, 356 (2008).

\bibitem{Gass08} M. H. Gass, U. Bangert, A. L. Bleloch, P. Wang,
R. R., Nair, and A. K. Geim, 
Nature Nanotechnology \textbf{3}, 676 (2008).

\bibitem{Kosynkin09} D. V. Kosynkin, A. L. Higginbotham, A. Sinitskii,
J. R. Lomeda, A. Dimiev, B. K. Price and J. M. Tour, 
Nature \textbf{458}, 872 (2009).

\bibitem{Liying09} L. Liying Jiao, L. Zhang, X., Wang, G., Diankov
and H. Dai, 
Nature \textbf{458}, 877 (2009).

\bibitem{Wang09} H. Wang, D. Nezich, J. Kong and T. Palacios, 
Electron Device Letters IEEE \textbf{30},547 (2009).

\bibitem{Schedin07} F. Schedin, A. K. Geim, D. V., Morozov, E. W.,
Hill, P. Blake, M. I. Katsnelson and K. S.Novoselov, 
Nature Materials \textbf{6}, 652 (2007).

\bibitem{Tour05} J. M. Tour, Molecular Electronics, 
(World Scientific Publishing, Singapore, 2005).

\bibitem{Bogani08} L. Bogani and W. Wernsdorfer, 
Nature Materials \textbf{7}, 179 (2008).

\bibitem{Dresselhaus-capulla} X. Jia1, M. Hofmann, V. Meunier, B.
Sumpter, J. Campos-Delgado, J. M. Romo-Herrera, H. Son, Y-P. Hsieh,
A. Reina, J. Kong, M. Terrones and M. S. Dresselhaus, 
Science, \textbf{323}, 1701 (2009).

\bibitem{Prins11} F. Prins, A. Barreiro, J. W. ruitenberg, J. S.
Seldenthuis, N. Aliaga-Alcalde, L. M. K. Vandersypen and H. S. J.
van der Zant, 
Nano Letters \textbf{11}, 4607 (2011).

\bibitem{He10} Y. D. He, H. L. Dong, T. Li, C. L. Wang, W. Shao,
Y. J. Zhang, L. Jiang and W. P. Hu, Appl. Phys. Lett. \textbf{97},
133301 (2010).

\bibitem{Cai10} J. Cai, et al., 
Nature \textbf{466}, 470 (2010).













\bibitem{Tour99} J. Chen, M. A. Reed, A. M. Rawlett and J. M. Tour,
Science \textbf{286}, 1550 (1999).

\bibitem{Park09} S. Barraza-Lopez, K. Park, V. M. Garc\'{\i}a-Suárez and
J. Ferrer, 
Phys. Rev. Lett, \textbf{102}, 246801 (2009).

\bibitem{Finch09}
C. M. Finch, V. M. Garc\'{\i}a-Suárez, and C. J. Lambert, Phys. Rev. B {\bf 79}, 033405 (2009).

\bibitem{Dragoman07} D. Dragoman and M. Dragoman, 
App. Phys. Lett. \textbf{90}, 143111 (2007).

\bibitem{Cheraghchi08} H. Cheraghchi and K. Esfarjani, 
Phys. Rev. B \textbf{78}, 085123 (2008).

\bibitem{Agapito07} L. A. Agapito and H. P. Cheng, 
J. Phys. Chem. C: \textbf{111}, 14266 (2007).

\bibitem{Koleini07} M. Koleini, M. Paulsson and M. Brandbyge, 
Phys. Rev. Lett. \textbf{98}, 197202 (2007)

\bibitem{Yokoyama08} T. Yokoyama, 
Phys. Rev. B \textbf{77}, 073413 (2008).

\bibitem{Cohen06} Cohen et al., Nature (2006).

\bibitem{Palacios07} J. Fernández-Rossier and J. J. Palacios, 
Phys. Rev. Lett. \textbf{99}, 177204 (2007).

\bibitem{Yazyev08} O. V. Yazyev and M. I. Katsnelson, 
Phys. Rev. Lett. \textbf{100}, 047209.

\bibitem{Eisele98} H. Eisele and G. I. Haddad, Modern Semiconductor
Device Physics, edited by S. M. Sze, Wiley, New York, 1998.

\bibitem{silicon} Titash Rakshit, Geng-Chiau Liang, Avik W. Ghosh,
and Supriyo Datta, 
Nano Lett. \textbf{4}, 1803 (2005).

\bibitem{Lyo89} I.-W. Lyo and Ph. Avouris, Science \textbf{245},
1369 (1989).

\bibitem{Dalgleish06} H. Dalgleish and G. Kirczenow, Phys. Rev. B
\textbf{73}, 245431 (2006).

\bibitem{Gar08} V. M. Garc\'{\i}a-Suárez and C. J. Lambert, Nanotechnology
\textbf{19}, 455203 (2008).

\bibitem{Chen07} L. Chen, Z. Hu, A. Zhao, B. Wang, Y. Luo, J. Yang
and J. G. Hou, Phys. Rev. Lett. \textbf{99}, 146803 (2007).







\bibitem{Soler02} J. M. Soler, E. Artacho, J. D. Gale, A. Garc\'{i}a,
J. Junquera, P. Ordejón, and D. Sánchez-Portal, J. Phys.: Condens.
Matter \textbf{14}, 2745 (2002).

\bibitem{LDA} D. M. Ceperley and B. J. Alder , Phys.\ Rev.\ Lett.\ \textbf{45},
566 (1980).

\bibitem{Roman09} G. Román-Pérez and J. M. Soler, 
Phys. Rev. Lett. \textbf{103}, 096102 (2009).

\bibitem{SMEAGOL} A. R. Rocha, V. M. Garc\'{i}a-Suárez, S. W. Bailey,
C. J. Lambert, J. Ferrer and S. Sanvito, Phys. Rev. B\textbf{73},
085414 (2006).

\bibitem{Dresselhaus-nr} K. Nakada, M. Fujita, G. Dresselhaus and
M. S. Dresselhaus, 
Phys. Rev. B \textbf{54}, 17954 (1996)

\bibitem{Adams12}
D. J. Adams, O. Gröning, C. A. Pignedoli, P. Ruffieux,
R. Fasel and D. Passerone, 
arXiv:1201.4735v1

\bibitem{Ferrer09}
J. Ferrer and V. M. Garc\'{\i}a-Suárez, Phys. Rev. B {\bf 80}, 085426 (2009).

\end{thebibliography}
\end{document}